\documentclass[12pt]{article}

\begin{document}

\def\eg{{\it e.g.}}
\def\etal{{\it et al.}}
\def\ni{\noindent}

\centerline{\bf \Large Astronomical Software Wants To Be Free: A Manifesto}

\begin{center}



\bigskip

a position paper for the 2010 Decadal Survey

submitted to the State of the Profession study groups on
Computation, Simulation and Data Handling, and Facilities, Funding and 
Programs

\bigskip

March 15, 2009

\bigskip

Benjamin J. Weiner\footnote{bjw@as.arizona.edu} (Steward Observatory),
Michael R. Blanton (NYU), Alison L. Coil (UCSD), Michael C. Cooper (Steward),
Romeel Dav\'e (Steward), David W. Hogg (NYU), Bradford P. Holden (UCO/Lick),
Patrik Jonsson (UCO/Lick), Susan A. Kassin (Oxford), 
Jennifer M. Lotz (NOAO), John Moustakas (NYU),
Jeffrey A. Newman (Pittsburgh), J.X. Prochaska (UCO/Lick),
Peter J. Teuben (Maryland), Christy A. Tremonti (MPIA/Wisconsin), 
Christopher N.A. Willmer (Steward)

\end{center}

\section{Summary}

Astronomical software is now a fact of daily life for all
hands-on members of the astronomy and astrophysics community.
Purpose-built software to assist in and automate data reduction 
and modeling tasks becomes ever more critical as we handle
larger amounts of data and simulations and doing steps ``by hand'' becomes 
less practical.  However, the writing of astronomical software is
unglamorous, the rewards are not always clear, and there are
structural disincentives to releasing software publicly and to
embedding it in the scientific literature,
which can lead to significant duplication of effort and
an incomplete scientific record.

In this position paper submitted to the 2010 Decadal Survey,
we identify some of these structural disincentives and 
suggest a variety of approaches to address them, with the
goals of raising the quality of astronomical software, 
improving the lot of scientist-authors, and providing
benefits to the entire community, analogous to the
benefits provided by open access to large survey and
simulation datasets.  Our aim is to open a conversation
on how to move forward.

We advocate that: (1) the astronomical community consider software as an
integral and fundable part of facility construction and
science programs; (2) that software release be considered as integral 
to the open and reproducible scientific process as are publication
and data release; (3) that we adopt technologies and
repositories for releasing and collaboration on software that 
have worked for open-source software; (4) that we
seek structural incentives to make the release of software 
and related publications easier for scientist-authors; (5) that
we consider new ways of funding the development of grass-roots
software; (6) and that we rethink our values to acknowledge
that astronomical software development is not just a 
technical endeavor, but a fundamental part of our scientific practice.

\section{The Landscape of Astronomical Software}

The need for purpose built software in the astronomical
enterprise, in the operation of telescopes and instruments,
data reduction, and the theoretical and modeling process,
is self-evident.  Astronomical software is generally 
custom written by scientists or by programmers hired
directly by scientific institutions, and covers a wide
range in both complexity and generality, including programs
written to solve a single problem; large pipelines
devoted to a specific instrument; and general purpose
subroutine libraries, data reduction, and simulation software.
However, many codes have to solve similar problems,
which is one reason that general subroutine libraries are
useful and that many astronomers build up personal
libraries of code whose pieces they use and reuse.
It is difficult to quantify the total amount of software or to
break out its expense, so this paper is admittedly anecdote
and opinion; we believe that many in the field will recognize
common experiences.

Some publicly released software packages have become industry
standards: examples include full data reduction suites 
(e.g. IRAF, AIPS, MIRIAD\footnote{Locations of software packages cited may
be found in the References.}) often involving teams of scientists and
programmers, 
and standalone packages frequently written by one or a few astronomers,
such as the DAOPHOT, SExtractor and GALFIT photometry and
analysis packages (Stetson 1987;
Bertin \& Arnouts 1996; Peng \etal\ 2002);
Bruzual and Charlot, P\'EGASE, and Starburst99 spectral synthesis
codes (Bruzual \& Charlot 2003; Fioc \& Rocca-Volmerange 1997;
Leitherer \etal\ 1999);
and simulation codes such as the Barnes-Hut treecode, FLASH, and
GADGET (Barnes \& Hut 1986; Teuben 1995; Fryxell \etal\ 2000; Springel 2005).  
There are also libraries that are widely used: e.g.
FITSIO, PGPLOT, the IDL Astronomy Library, IDLUTILS.\footnote{Many, 
though not all, of the packages and libraries named
here have been written or maintained by staff astronomers at
observatories or national laboratories, providing a service
to the whole community.}
And there are data reduction packages
that are public and/or have been adapted for multiple instruments.
The advantages of common and public software are twofold: an
enormous amount of time and duplication of effort is saved by 
using existing reliable packages\footnote{As a concrete example,
the spectroscopic pipeline for DEIMOS and the DEEP2 survey 
(Davis \etal\ 2003) relied heavily on the donation of much material 
from an SDSS spectroscopic pipeline (http://spectro.princeton.edu), 
even though they are very different datasets.  The cores of each
of these pipelines, written by close-knit groups of just a few talented
astronomers, took several person-years to bring to version 1.0.}, 
and because standard methods are used, results across different 
papers/projects are more easily compared.

In the opinion of the authors, useful public software packages
written by lone astronomers or small teams, such as
DAOPHOT and SExtractor, have enabled easily as much science as 
yet another large telescope would have, at considerably lower cost.
However, the community is dependent on the goodwill and industry of
a small number of authors to release and maintain such packages.
Beyond writing a package, the effort to document, release, and 
provide even minimal support is substantial.

For every program that is released, there are
many useful, if less fully developed, programs and subroutines 
languishing on people's disks or even orphaned, and 
similar codes get written over and 
over again.  There is a large amount of experience, knowledge and sunk
costs going to waste because our methods, motivations and rewards for 
sharing code, collaborating on it, making public releases,
and giving credit are even less well developed than the methods,
motivations and rewards for sharing data and making data public.
Many observing calls and funding programs require or
encourage that the investigators make periodic data releases,
there are repositories to collect such data, and there are
protocols for citing and acknowledging the use of data releases,
but there are few analogous or widely-used mechanisms for software.

\section{Software as Last Line in the Budget}

The great property of software is that the entire community 
speaks the language: nearly everyone with an astronomer's training
has written at least some code, although abilities and inclinations
vary widely.  The unfortunate corollary is that
familiarity breeds contempt: software is often regarded
as a responsibility that can be handed off to anyone without requiring 
explicit planning or much funding and support.  This leads
to both a tendency not to fund software, and a tendency not
to regard highly productive scientist-programmers as elite
scientists.

For example, astronomical 
projects to build multi-million dollar instruments generally 
contract for specialized mechanical, optical, and electrical
engineering, but often then hand the end product of the data over
to an ad hoc team of astronomers, frequently junior personnel,
to patch together reduction
software.  Because reduction software can realistically only
be finished with real data in hand, it is the final step and
the most squeezed by cost overruns and time delays.  Even
projects with the best intentions find themselves running out
of time and money to complete pipeline software that is robust 
and does not require fine-tuning, and the release
of a polished package to the user community is frequently
long delayed.  

The resulting cost of wasted resources is substantial.
Many observers have optimistically taken data that
they have never managed to publish because the data reduction
experience is too complex or the software tools are inadequate, 
perhaps because the instrument team was never
funded to produce and release a pipeline, or because the tools
that were produced are not robust enough to handle all
data-taking modes.  Again, this often happens despite the good
intentions of the developers.  
Once an instrument is on the sky, a survey is 
under way, or a simulation
code is written, the project team is under pressure to 
``do science'' (a phrase which reflects our community
bias that instruments and software are not science) and put 
aside bomb-proofing and releasing code.

We emphasize that for most projects {\it there is nothing wrong with
relying on small teams of scientists to write software}.  The
problems are rather that this activity is generally not planned
or adequately funded, and that the community does not offer 
adequate reward or incentive for having done it well and releasing
software products publicly.  When adequate inducements,
funding, or pressure have been applied, small groups have
turned out excellent public instrument data reduction, analysis,
and theory packages.

We do not argue that small groups of often-junior scientists
ought to be replaced by teams, large or small, of dedicated 
programmers, either scientists or software engineers.  This
paper is concerned with aiding grass-roots software efforts,
complementary to top-down or standards-definition efforts such
as the National Virtual Observatory.
The problems of project management with an increased number of 
people\footnote{The ``mythical man-month'': ``Adding manpower
to a late software project makes it later'' (Brooks 1975).} and of 
communication between scientist users and programming team
members are well known.  Large teams of programmers can be
appropriate for observatory-class missions, but these are
outside the scope of the issues we address in this paper.
There are examples in the astronomy community of software challenges
that have been solved more effectively by small groups of 
scientists than by large teams.  
Our goal should be to
foster the conditions under which scientists can bring
software projects, small or large, to fruition and as
a critical stage, to release code so that it benefits
and can be reused by the community, and to recognize
the achievements of the authors.

\section{Disincentives for Programming and Software Release}

The problem of astronomical software as common language,
in which many are fluent, is that {\it because
software can seemingly be done cheaply, it is frequently done on the cheap}.
This means that its creators frequently do not have time
to truly finish it, much less document and release it, before the
demands of moving on to ``do science,''  or to move on to
finishing grad school or a postdoc or building the next project.
In many ways, software
work suffers from similar problems to those that instrument builders
face, with the demand to ``do science'' after one has
spent a long time building an instrument or facility.
However, software work is even more often discounted since
it does not have a physical, tangible product. 

Even when a program is largely complete, the additional
effort of releasing
it, documenting it, and especially attempting to write it
up for publication is time consuming.  The
documentation and release webpage are tedious
to maintain and frequently suffer from ``link rot,''
especially for junior people who change locations.
These problems are recognized and ongoing for astronomical
data releases; a number of observatories maintain data
archives and there are catalog repositories at the
ADS, CADC, and CDS, but we do not have a similarly
transparent system for software releases.

Although a few
algorithms and packages become industry standards and
articles describing them garner many citations
(e.g. Stetson 1987; Bertin \& Arnouts 1996), publishing
methods articles is generally not a good bet.  A junior
person who develops software such as a pipeline 
is well advised not to
publish on it, but to write science-result papers instead,
reflecting the current value system of the community.
Software releases also rarely
garner wide recognition, and without an accompanying paper to cite,
the work cannot be paid back in citation currency.  

Additionally, emphasizing one's work on software or
publishing methods papers carries a risk of being perceived
as a programmer first and scientist second, falling on the
wrong side of the technician/scientist divide.  We believe
this is a false dichotomy, but especially in an age
of large projects, it is an increasing sociological
problem within the community.

Recognizing the disincentives to software work and its release
as a problem and addressing it 
does not mean throwing money at it
by increasing the size of programming teams.  It can mean 
providing the existing people more time, or more resources to 
explicitly encourage and fund
the production of public software releases, just as Legacy/Treasury
observing programs mandate and fund the production of
public data releases.  It may require devoting resources
to retain the most talented and productive scientist-programmers,
and to training young scientists to write good software, rather
that expecting them to pick it up as they go along.
It suggests the more difficult task
of reconfiguring our vision of ``doing science'' to include
software as an integral part.

The facts that software is a {\it lingua franca},
that so many of us can work on software and
that talented people can emerge without requiring outside
expertise are not reasons to take software efforts for
granted, but rather argue for more openness to foster 
collaboration, testing, maintenance and improvement of
astronomical software packages.


\section{What Is To Be Done? Creating Incentives to Do Better}

At a fundamental level, the astronomical community needs to
change its culture to reflect that we are software dependent and 
that creation, release and maintenance of astronomical software 
is an integral and valuable part of our enterprise.  However,
changing our culture is neither easy nor a concrete recommendation.
We offer several suggestions for steps toward improvement,
which vary in expense and commitment.  Our aim is to get the community
thinking about new or innovative solutions.

\medskip

\ni{\bf Suggestions for moving forward:}

\medskip

1. Borrowing from technology 
developed in the
open source movement, we should {\bf create a open central 
repository location at which authors can release software and 
documentation} by uploading it, with version control, and license
it to be modified by other users.  Users could
search for software, submit comments, bug reports and fixes, 
ask and answer questions and add to the documentation.
This will reduce the burden on each author of maintaining 
individual software release pages, versioning and bug tracking systems, 
and make it easier for users to search for appropriate software.  

User-generated documentation, ideally in wiki format,
and identification of problems,
would add considerable value.  Examples of such content already
exist, such as webpages written by individual astronomers
that are introductions or tutorials for packages or
instruments, e.g. IDL/IRAF/SM/data reduction.
A forum in which users can ask and answer questions (analogous
to those at iraf.net or terapix.iap.fr) would allow for 
easier maintenance and reduce the burden of duplicate
queries to authors.  Several packages of open-source
software to run repositories, wikis and forums exist now, e.g. 
the sourceforge.net repository.  What we advocate is a dedicated
version for astronomy that would be attractive enough to become
commonly used within the community: it would not be very expensive, and 
it's only surprising that it doesn't already exist.

2. {\bf Software release should be an integral and funded part of astronomical 
projects.}  Many surveys have data release requirements from the
funding entity, e.g. observatory Legacy surveys and some NSF-funded surveys.  
Software release should also be a
deliverable (and thus a fundable expense).  NSF-funded instruments should
budget for and provide a reduction pipeline, open-sourced
to the entire community, not restricted to the users of any
single facility.  Proposals of all types, including those for surveys 
and theory programs, should consider releasing
software to make their science products easier to use, and
this should be a proposal grading criterion.

3. {\bf Software release should become an
integral part of the publication process.}
A key element of scientific publication is that the 
results are in principle reproducible.  As software becomes
too complex to describe fully in the methods section of a 
paper, software release will become a more urgent issue.
The community should consider how best to link publications
to released electronic products; printing web addresses in
papers is prone to link rot.  
Formal requirements for release are probably unenforceable; there
will always be situations such as grad students
with theses in progress to protect.  However, we believe that
people who release software will find that it has a greater
impact, will gain collaborators, and that this will outweigh
concerns about losing a proprietary advantage over their
competitors.  Several of the software packages cited earlier
bear this out. In general, the most efficient way
to take advantage of someone's software is to collaborate with
the author.  We see software releases as an avenue to encourage
such collaborations.

4. {\bf The barriers to publication of methods and 
descriptive papers should be lower.}  In large part this is a cultural
issue, as it is already possible to publish
methods papers and short research notes in journals
such as PASP.  We should encourage the publication of
methods papers, consider subsidizing their publication
through small grants, and find ways to expedite the writing
and refereeing process.

5. {\bf Astronomical programming, statistics and data
analysis should be an integral
part of the curriculum} for undergrad and grad students,
using contemporary techniques.  This reflects its place
as an inescapable part of our scientific practice.  
Many of today's students arrive with vastly different computer
experiences than we had in previous generations, and
training is key to producing software-literate scientists.
Training should adapt to today's needs; for example,
in an era of large datasets, students must master the
use of data structures, and training that is solely 
number-crunching will not serve them well.

6. We should  {\bf encourage interdisciplinary cooperation} with
like-minded and algorithmically sophisticated members
of the computer science community, for both research
and education/outreach purposes.  Examples of this 
in practice include the Astrometry.net project and
Google Sky.

7. We should create {\bf 
more opportunities to fund grass-roots software projects
of use to the wider community,}
whether through existing grants programs or new calls.
Currently, software, archival analysis, and writing and 
releasing code to make datasets more easily usable can
be difficult to fund through normal channels, as they
lack the flash of the new.  Especially in an era
of large surveys, analysis projects will take on greater
importance.  The current situation is heading for an impasse,
where large archives and datasets are mandated for release
but it may be difficult to fund projects to work with them;
astronomers will rather try to obtain new observations if those
come with analysis funding.

8. We should develop {\bf institutional support for science 
programs that attract and support talented scientists who generate 
software for public release.}  These
could include: individual institutions recruiting faculty 
who can exploit the ever-higher tide of public data
(especially useful for institutions that have not bought
into expensive large telescopes);
collaborations on problems of interest to the
community, along the lines of NSF Grand Challenges;
proposals for software centers, potentially along the lines
of the CADC or of theory shops such as the IAS or CITA, but with a less
strictly-theory mission; or an NSF Science and Technology
Center devoted to astrophysical inference, statistics and
analysis.  Centers would have to be carefully managed
to be compatible with our grass-roots, bottom up model 
of encouraging widespread software development.
However, providing a stable career path for scientists
with these skills would benefit a wide range of 
projects and future surveys.


\section{Conclusions}

We argue that the astronomical community's increasing dependence
on software and analysis requires that it come to see software
as an integral part of its scientific practice.  Elevating the 
status of software does not mean hiring a new army of programmers,
but recognizing the achievements of the scientists who
create the tools we use, and removing barriers to the 
production, completion, and sharing of astronomical
software.  We advocate greater openness in software
release and incentives for this, borrowing from the
techniques that have made open source software successful.

Inevitably some software will be resoundingly useful on 
first release, some will go through generations of improvement,
some will be used in its original version without updates or
maintenance, and some will be released and never used.
This is a perfectly fine outcome.  A software ecosystem
will let the community decide what is
most useful and develop collaborations 
through an evolutionary process, rather than
acting as gatekeepers or keeping code proprietary for
transitory advantages.  The process of testing, comparison,
challenge, and incremental progress is at the heart of
free scientific inquiry, and it can free our code as well.

\bigskip

{\parindent 0pt
{\bf References:}

\bigskip


Barnes, J., \& Hut, P.\ 1986, Nature, 324, 446

Bertin, E., \& Arnouts, S.\ 1996, A\&AS, 117, 393

Brooks, F. 1975, {\it The Mythical Man-Month}, Addison-Wesley

Bruzual, G., \& Charlot, S.\ 2003, MNRAS, 344, 1000

Davis, M. \etal\ 2003, Proc. SPIE, 4834, 161

Fioc, M., \& Rocca-Volmerange, B.\ 1997, A\&A, 326, 950

Fryxell, B. \etal\ 2000, ApJS, 131, 273

Leitherer, C., et al.\ 1999, ApJS, 123, 3

Peng, C.Y., Ho, L.C., Impey, C.D., \& Rix, H.-W.\ 2002, AJ, 124, 266 

Springel, V.\ 2005, MNRAS, 364, 1105 

Stetson, P.~B.\ 1987, PASP, 99, 191 

Teuben, P.J. 1995, ADASS IV, ed. R. Shaw \etal, PASP Conf. Series 77, p. 398


\bigskip

Software packages referred to:

\bigskip

AIPS: http://www.aips.nrao.edu/  \quad AIPS is distributed by NRAO; radio reduction software

Astrometry.net: http://astrometry.net \quad  Public astrometric calibration

DAOPHOT: Stetson 1987, included in several major software distributions \quad Crowded-field stellar photometry

DEEP2/DEIMOS Public Pipeline: http://astro.berkeley.edu/$\sim$cooper/deep/spec2d/ \quad  Public spectroscopic data reduction

FITSIO: http://heasarc.gsfc.nasa.gov/fitsio/ \quad FITS file interface library

FLASH: http://flash.uchicago.edu/ \quad Modeling code for thermonuclear flashes, developed collaboratively 

GALAXEV, Bruzual \& Charlot modeling code: http://www.cida.ve/$\sim$bruzual/bc2003

GALFIT: http://users.ociw.edu/peng/work/galfit/galfit.html \quad Galaxy photometric model fitting

IDL Astronomy Users' Library: http://idlastro.gsfc.nasa.gov/ \quad Broad range of library routines

IDLUTILS: http://spectro.princeton.edu/idlspec2d\_install.html \quad IDL utilities

IRAF: http://www.iraf.net/ \quad  IRAF is distributed by NOAO; data reduction software and user forum

MIRIAD: http://bima.astro.umd.edu/miriad/ \quad Radio reduction software

NEMO, including the Barnes-Hut treecode: http://www.astro.umd.edu/nemo/ \quad Stellar dynamics codes

P\'EGASE: http://www2.iap.fr/users/fioc/PEGASE.html \quad Spectral synthesis models

PGPLOT: http://www.astro.caltech.edu/$\sim$tjp/pgplot/ \quad Plotting utilities

SDSS Pipeline: http://spectro.princeton.edu/ \quad Public spectroscopic reductions

SExtractor: http://terapix.iap.fr/rubrique.php?id\_rubrique=91 \quad Faint galaxy photometry

Starburst99: http://www.stsci.edu/science/starburst99/ \quad  Spectral synthesis models
}

\bigskip

\quad\bigskip

\noindent Endnote: The title paraphrases Stewart Brand's influential,
though highly ambiguous, maxim that ``Information wants to be free.''  
Free to use, as in open-source code, or free as in no one is willing 
to pay its true cost?  

\end{document}